# Evidence of Dark Energy Prior to its Discovery

Geoffrey W. Marcy

Center for Space Laser Awareness, USA

## Abstract

This is a review and statistical analysis of the evidence supporting the existence of a cosmological constant in the early 1990s, before its discovery made with distant supernovae in 1998. The earlier evidence was derived from new and more accurate measurements of the Universe, including its mass density, the Hubble constant, the age of the oldest stars, the filamentary large-scale structure, and the anisotropy of the cosmic microwave background. These measurements created tension for models that assumed the cosmological constant was zero. This tension was alleviated by several insightful papers published before 1996, which proposed a cosmological constant that increased the expansion rate. Statistical analysis here shows that the probability of the cosmological constant being zero was demonstrably less than a few percent. Some models identified a best-fit value close to the modern estimate of $\Omega_\Lambda \sim 0.7$.

## 1. Introduction

Until 1998, most theories describing the structure of the Universe focused on its expansion rate, its density, and standard general relativity. They rarely considered a cosmological constant or any form of "dark energy."[1] This exclusion of the cosmological constant was justified because there was no direct evidence for its existence, and some theoretical predictions found "the Cosmological Constant is probably zero."[2] In fact, during the mid-1990's, research teams calibrating Type Ia supernovae were primarily concerned with determining the Hubble constant and the *deceleration* of the Universe's expansion, not with investigating any cosmological constant.[3]

In the early 1990's, remarkable progress was made in obtaining precise measurements of various properties of the Universe. These properties included its mass density, the Hubble constant, the age of the oldest stars, its large-scale structure (including dark matter), the large-scale anisotropy of the cosmic microwave background (CMB), and the "horizon problem". These measurements were conducted by different teams, independently of each other and without the influence of existing theories, minimizing the risk of cross-contamination. Some theorists observed that these independent observations were mutually inconsistent, which led to the exploration of solutions involving a cosmological constant.

However, the cosmological constant has a complex and controversial history. Einstein initially introduced it into his general relativity field equations to prevent the prediction that the Universe's mass would cause space to contract over time due to gravity. However, Vesto Slipher's long-exposure photographic spectra (taking more than 24 hours) revealed that most "spiral nebulae" exhibited redshifts, indicating that they were moving away from us at speeds exceeding 300 km/s.[4]

Seeing these measurements, Lemaître brilliantly noted a linear relation between distance and redshift, and he rejected the concept of a static universe.[5] He proceeded to use general relativity to derive a theory of an expanding Universe having constant mass, where recession velocities would be proportional to distance. Recognition for this discovery came 89 years later when the International Astronomical Union renamed the proportional relationship the "Hubble-Lemaître Law". Revisiting and correcting historical narratives provide valuable insights into the development of scientific theories. Similarly, this paper offers a suggestion that diverse observations can be assembled by thoughtful theorists to offer insight about heretofore undetected components of the Universe.

Lemaître strongly advocated for the cosmology community to retain the cosmological constant.[6] His Ph.D. thesis included a mathematical analysis demonstrating that the most general mathematical formulation of Einstein's field equations required inclusion of a cosmological constant.[7] He quickly realized that the expansion of the Universe implied a microscopic origin - a "primeval atom", later termed the "big bang."[8]

This insight led him to explore the time-dependent expansion history of the Universe. He noted that a slow expansion followed by acceleration rendered the universe old enough to form stars and galaxies, while satisfying the current expansion rate defined by the Hubble constant.[9] His quantitative analyses of cosmological evolution, predictions of a cosmic microwave background and galaxy evolution, and the need for temporal self-consistency, including measured ages of meteorites from radioactive Uranium, remain brilliant and relevant.[10] Lemaître understood that his computed expansion age of the Universe was too young to provide time to form galaxies, but was ameliorated by invoking a cosmological constant. Lemaître's issue of temporal self-consistency in the expansion history of the Universe is the conceptual essence of the research reviewed here. Nonetheless, although the possibility of a non-zero cosmological constant was sometimes included in theory, it was not supported by empirical evidence and indeed Einstein himself discounted it as *ad hoc*.

Abandoning the cosmological constant, most astronomers spent 70 years attentive to two key properties of the Universe, the rate of its expansion, $H_o$, and the *deceleration* of that expansion due to the mutual gravitational pull of its galaxies. They also made extensive measurements of the overall mass density to determine whether the Universe would eventually recollapse due to gravity or continue expanding forever. These measurements led to the discovery of dark matter, which was detectable through various methods, including from rotation curves of galaxies, the velocities of galaxies bound within clusters, gravitational lensing around galaxy clusters, and x-ray emissions from hot gas in clusters.[11] In the 1930s, Fritz Zwicky first proposed the existence of dark matter after observing that the Coma Cluster's galaxies moved too fast for their visible mass to account for the gravitational pull holding them together. Vera Rubin and Kent Ford provided further evidence in the late 1970s by measuring the rotation curves of spiral galaxies, finding that stars at the edges orbited at unexpectedly high speeds, implying the presence of unseen mass.

By the early1990's, measurements indicated that the combined density of dark and ordinary matter was only 10% to 20% of the amount needed to cause the Universe to recollapse, corresponding to a matter density parameter ($\Omega_M$) of 0.1 to 0.2. The modern value of $\Omega_M$ is now known to be ~0.3, but even the earlier value of 0.1 to 0.2 already suggested that our Universe has insufficient mass on its own to make it "flat", meaning it lacks enough mass to halt its expansion.

By 1990, three major conflicts emerged for a Universe composed only of cold dark matter (CDM) and a small amount of ordinary matter. First, the measured ages of the oldest stars, ranging from 13 to 16 billion years, were older than the calculated expansion time since the Big Bang, which was about 10.5 billion years – an impossibility (described below). This age discrepancy was evident by 1993 and worsened by 1995 due to new measurements of stellar ages and a Hubble constant of $H_o \approx 77$ km s$^{-1}$ Mpc$^{-1}$. The implied expansion age of the Universe, calculated without a cosmological constant, was 10.8 billion years, making it younger than the oldest stars, clearly an absurd result.

Second, simulations of galaxy formation and of the large-scale structure of the Universe struggled to reproduce the observed large-scale structure and were often able to achieve a good fit by introducing a cosmological constant, as described below.[12] One challenge was simulating the gravitational collapse of the primordial mass clumps into the filamentary structure observed within the age of the Universe.[13]

Third, a challenge arose in explaining the near uniformity of both the CMB and large-scale structure of the Universe in all directions, known as the "horizon problem". Opposite regions of the Universe are not in causal contact, meaning they cannot exchange information or equilibrate, yet they are remarkably similar. A proposed theoretical solution came from Alexei Starobinskii, Alan Guth, Andrei Linde, Andy Albrecht and Paul Steinhardt, who developed the concept of post-Big Bang "inflation."[14] This idea suggests that an exponential expansion in the early Universe allowed for prior causal connections within the observable Universe. These models predicted that the Universe should be nearly "flat". However, the measured mass density at the time was (and is today) too low to cause a flat geometry by itself.

These multiple conflicts motivated reconsideration of the cosmological constant.[15] For example, Turner published that the current measurements, "…point to a best-fit model with the following parameters: $\Omega_{baryon} \approx 0.03$, $\Omega_{dark\ matter} \approx 0.17$, $\Omega_\Lambda \approx 0.8$, and $H_0 \approx 70$ km/sec/Megaparsec, which improves significantly the concordance with observations."[16] Krauss and Turner wrote, "Unless at least two of the fundamental observations described here are incorrect a cosmological constant is required by the data."[17]  These papers offered compelling reasons to invoke a cosmological constant in the early 1990's, well before the supernova results.

The following sections present the evidence in succession. The next section presents a quick summary of the relevant general relativity that governs the expansion rate. The following section presents the newly obtained measurements of the Universe by 1995.  The next section reviews the remarkable papers that describe the tension that resulted from the new measurements, motivating inclusion of a cosmological constant.  The next sections present the statistical probability, by 1995, that a significant cosmological constant was favored, even when employing modern properties of the Universe.

## 2.  The Friedmann Equation and Expansion with Time

To review, Einstein's field equations combined with the Friedmann-Robertson-Walker metric lead directly to the Friedmann Equation, which was also derived by Lemaître.[18]  This equation provides an expression for the expansion of the Universe over time[19]:

$$H^2 = \left(\frac{\dot{a}}{a}\right)^2 = \frac{8\pi G}{3}\rho_M - k/a^2 + \Lambda/3 \qquad (1)$$

Here $H$ represents the Hubble "constant", which is actually a function of time.  The variable $a$ is the scale factor representing the size of the Universe.  The term $\rho_M$ denotes the large-scale mass density of the Universe, including any contribution from radiation.  The parameter $k$ represents the curvature of space, with $k$ equal to 0 indicating a flat Universe.  Finally, $\Lambda$ is Einstein's cosmological constant, measured in units of meters$^{-2}$ for c=1.

Commonly, one encapsulates the key components of the Universe as,

$$\Omega_M + \Omega_\Lambda = \Omega_{tot} \qquad (2)$$

Here, $\Omega_M$ and $\Omega_\Lambda$ represent the fractions of the total mass-energy density of the Universe contributed by mass and "dark energy", respectively, normalized to a flat Universe where $\Omega_{tot} = 1$. A value of $\Omega_{tot}$ less than 1 indicates an open Universe, while a value greater than 1 indicates a closed geometry.  By 1993, observational constraints suggested that $\Omega_M$ was 0.1 to 0.3 (see the following section), and $\Omega_\Lambda$ was generally assumed to be zero or ignored, leading to an estimate of $\Omega_{tot} \approx 0.2$.  This implied an open Universe, destined to expand forever.

Yet, the plausibly flat geometry implied by theory, especially to solve the horizon problem,[20] predicted that

$$\Omega_M + \Omega_\Lambda = 1 . \qquad (3)$$

One may integrate eq. 1 analytically or numerically to compute the time since the Big Bang, the time since $a = 0$, as a function of $H_o$, $\rho_M$ and $\Lambda$, or equivalently as a function of $H_o$, $\Omega_M$ and $\Omega_\Lambda$.[21] Carroll et al. provide useful approximations for the age of the Universe as a function of $H_o$, $\Omega_M$ and $\Omega_\Lambda$.[22]

Extensive efforts were made to measure the deceleration of the Universe using standard candles, such as Cepheid variables, planetary nebulae, supergiants, horizontal branch stars, and both Type I and Type II supernovae. Additionally, other observables were measured to test the validity of a matter-only Universe that was decelerating. These observables included $H_o$, the density of the Universe, the ages of the oldest known stars, the large-scale structure of the Universe, and the anisotropy of the cosmic microwave background, as summarized in the section below.

## 3. Four Key Measurements of the Universe by 1995

By 1995, large teams had measured four crucial cosmological quantities with enough accuracy to impose strict constraints on the cosmological constant. In this review, those four measurements are described, along with a fifth section addressing indications of a flat universe.

**The Hubble Constant**
By 1992, the Hubble constant was measured independently using various standard candles. The main challenge in each method was accurately calibrating distances using luminous markers located many megaparsecs away. Redshifts were clearly determined through spectroscopic measurements of wavelength shifts of spectral lines. Jacoby *et al.* employed seven different distance indicators to galaxies, including the "standard candle" luminosities of planetary nebulae in galaxies within the Virgo and Coma clusters.[23] Their findings showed that the seven distance indicators yielded an average Hubble constant of $H_0 = 80 \pm 11$ and $73 \pm 11$ km s$^{-1}$ Mpc$^{-1}$, depending on the weighting scheme.

Van den Bergh offered another comprehensive analysis of the Hubble constant measurements, incorporating corrections for systematic errors in distance measurements to the Virgo and Coma clusters, resulting in $H_0 = 76 \pm 9$ km s$^{-1}$ Mpc$^{-1}$.[24] Van den Bergh noted the tension, "The observed value of $H_0$ differs at the 3-sigma level from that predicted by theories of stellar evolution in conjunction with canonical models of the Universe with $\Omega=1$ and $\Lambda = 0$."[25] ( By $\Omega$ he meant matter only.) This 3-sigma discrepancy between the age of stars and the Universe's expansion age served as an early warning against assuming $\Lambda = 0$. Freedman et al. and Kennicutt *et al.* made brilliant use of the Hubble Space Telescope within their "Key Project" to observe the pulsations of Cepheid variable stars as standard candles in distant galaxies, providing a preliminary value of the Hubble constant, $H_0 = 80 \pm 17$ km s$^{-1}$ Mpc$^{-1}$.[26]

*In summary, by 1995 three comprehensive papers reported four values for $H_0$ of 80, 73, 76, and 80 km s$^{-1}$ Mpc$^{-1}$, resulting in a weighted average of $77 \pm 6$ km s$^{-1}$ Mpc$^{-1}$.*

The Friedmann equation for a Universe with $H_0 = 77$ km s$^{-1}$ Mpc$^{-1}$ and $\Omega_M \sim 0.2$ (see the next section) but with no cosmological constant yields an expansion age of the Universe of 10.7 Gyr. In contrast, the age of the oldest stars known in 1995 was 12 to 16 Gyr (see the following two sections). The tension was emerging.

For completeness, Allan Sandage and collaborators wrote a series of more than 15 papers between 1970 and 2006, reporting measurements of $H_0$ mostly in the range, 45 to 62 km s$^{-1}$ Mpc$^{-1}$. However, by the mid-1990's the measurements by the Hubble Space Telescope, referenced above, showed that the new distance calibration was superior.

The most recent measurements of the Hubble constant are 73.0 ± 1.0 km s$^{-1}$ Mpc$^{-1}$,[27] 73.29 ± 0.90 km s$^{-1}$ Mpc$^{-1}$,[28] 71.76 ± 1.19 and 73.22± 1.28,[29] and 69 km s$^{-1}$ Mpc$^{-1}$.[30] The straight average of these is 72.0 ± 2. This value implies an expansion age of 11.0 Gyr, assuming no cosmological constant. Looking ahead, this expansion age calculated with a modern Hubble constant is less than the smallest likely age of the oldest stars known in 1995, 12 Gyr, indicating an inconsistency.

**The Mass Density of the Universe, $\Omega_M$**
By 1995, measurements of the Universe's total mass density, including both normal and dark matter within galaxies, the halos of dark matter surrounding them, and the hot, X-ray-emitting gas between the galaxies, had been conducted. Orbital rotation curves (Rubin *et al.* 1978; Rubin 1983; Rubin 1993) provided accurate mass measurements for the inner ~15 kiloparecs of galaxies.[31] Additionally, the velocity dispersion of galaxies within clusters and our infall velocity towards the Virgo cluster helped determine the total mass density on spatial scales up to 200 kpc from galaxies, resulting in a value of $\Omega_M$ = 0.2 to 0.3.[32] This value was in agreement with mass measurements on smaller spatial scales.[33]

Additional mass measurements on Mpc scales came from the x-rays emitted by the million-degree gas in rich galaxy clusters. The x-ray emission gave a direct measure of the amount of atomic (baryonic) matter throughout the cluster of galaxies, and more importantly the temperature of that gas offers a virial measure of the total mass in protons, neutrons, and dark matter. Briel *et al.* made such measurements with the ROSAT x-ray telescope, concluding, "Furthermore, if the matter in the Coma Cluster is representative of the Universe ... hence $\Omega_M < 0.17\ h_{50}^{-1/2}$."[34] Here, $h_{50} = H_0/50$, implying $\Omega_M < 0.14$.

Bahcall *et al.* reviewed dynamical and x-ray fluxes from galaxy clusters finding, "On cluster scales of ~ Mpc, the dynamical estimates of cluster masses − using optical, x-ray, and some gravitational lensing data − suggest low-densities: $\Omega_M \sim 0.2$"[35] and Bahcall, Fan, and Cen found $\Omega_M = 0.3 \pm 0.1$.[36] Excellent reviews of the mass density of the Universe are provided by Strauss and Willick, as well as by Davis, Nusser, and Willick.[37] The modern value that includes gravitational lensing measurements is $\Omega_M = 0.31 \pm 0.01$.[38]

**The Age of the Oldest Stars and Thus of the Universe**
The age of the stars in the oldest globular clusters, identified by their low abundance of iron and other heavy elements, sets a minimum age for the Universe. Many groups conducted photometry of these clusters and built stellar interior models to determine the best-fit ages, typically based on the main-sequence turnoff. Chaboyer *et al.* studied the 17 oldest globular clusters by performing Monte Carlo simulations of stellar isochrone fitting using color-magnitude diagrams.[39] They varied parameters to account for uncertainties in photometry, distances, and interior physics according to their probability distributions. Chaboyer *et al.* found a 95% confidence lower bound for the age distribution at 12.1 Gyr.[40] Bolte and Hogan estimated globular cluster ages at 15.8 ± 2.1 Gyr.[41] A review by VandenBerg, Bolte, and Stetson concluded, "Ages below 12 Gyr or above 20 Gyr appear to be highly unlikely."[42]

This 2σ lower limit for the age of the oldest clusters, ranging from 11.6 to 12.1 Gyr, will be used in later sections here to calculate the probabilities for the age of the Universe, posing a significant challenge to any model that excludes a cosmological constant. The reliability of these findings is compromised by any systematic errors common to all studies. The modern value for the ages of the oldest stars is ~12.3 Gyr (described below).

Additionally, the oldest white dwarfs, cooling at predictable rates, were found to be 10 Gyr old.[43] These must have spent at least 2 Gyr as normal hydrogen-burning stars before becoming white dwarfs, implying the Universe is at least 12 Gyr old. This estimate aligns with the 2σ lower limit of 11.6 Gyr derived from the ages of globular clusters. Therefore, the Universe was found to be older than 12 Gyr with about a 95% probability.

**Filamentary Large-Scale Structure and the Cosmic Microwave Background**
The groundbreaking CfA Redshift Survey, which included 3,000 galaxies up to a redshift of 0.04, revealed that galaxies are distributed in filaments and voids across scales of tens of millions of light-years.[44] Within a narrow angle of just a few degrees, the Universe exhibited variations in galaxy densities, forming a peculiar "soap-bubble" structure that demanded an explanation, particularly through simulations tracing back to fluctuations shortly after the Big Bang. At the same time, on larger angular scales, the Universe appeared uniform in all directions, also requiring explanation.

In a remarkable convergence of related data, new observations were made of the isotropy and superimposed temperature fluctuations of the cosmic microwave background (CMB). Uson and Wilkinson measured a stringent upper limit on the root-mean-square temperature fluctuations at a small angular scale of 4 arcminutes, finding a fractional fluctuation of less than $2.1 \times 10^{-5}$ against the 2.7 K background temperature of the CMB.[45] These CMB measurements placed an upper limit on the magnitude of the initial fluctuations that would later form galaxies and large-scale structures.

In 1994, the COBE experiment measured large-angular-scale (10°) fluctuations in the CMB, further supporting the presence of $10^{-5}$ fractional temperature fluctuations.[46] These spatial fluctuations in the CMB constrained the amplitude of the density fluctuations that eventually grew into the large-scale structures observed in the CfA survey, allowing simulations to check the consistency of cosmological models.

These CMB measurements also confirmed the coherence of the Universe on horizon scales as early as 380,000 years after the Big Bang. The fact that CMB radiation from opposite directions, just now reaching Earth, has the same temperature (2.7 K) suggests there was no opportunity for temperature equilibration, raising two significant challenges for cosmological models: explaining the fluctuation seeds of large-scale structure and addressing the horizon problem, which seems to defy causality. This led to the critical question of whether cold dark matter (along with small amounts of baryonic matter) was sufficient to account for these observations without invoking a cosmological constant.[47]

Davis *et al.* did not include a cosmological constant at all in simulating large scale structure and associated velocity fields.[48] Bahcall and Fan considered a cosmological constant in their simulations of large-scale structure, but found no significant difference in the simulated structure between two cases of an open Universe and one flat due to the inclusion of Λ.[49]

The high spatial-resolution (0.5 deg) COBE results that revealed the amplitude and angular scale of the fluctuations would not be available until after ~2000. Nonetheless, the upper limits on fluctuation amplitudes and the large-scale isotropy of the CMB already imposed an initial condition for simulations of the present-day large-scale structure. The simulations suggested that including a cosmological constant or not were equally consistent with observed large-scale structure.

**Semi-Empirical Evidence for a Flat Geometry**
Leading cosmologists argued that the Universe must be nearly flat to avoid two extreme outcomes: early recollapse or rapid, unchecked expansion.[50] In a nearly flat Universe, self-gravity within matter clumps can overcome the expansion of space, allowing sufficient time for galaxies to form and grow through gravitational attraction. The well-known "horizon problem" is related to flatness. The Universe appears isotropic on large scales at redshifts greater than 1.0, even though regions in opposite directions are not in causal contact with each other. This suggests an early epoch where causal equilibration was possible, followed by an inflationary period that pushed these regions out of causal contact. Inflation models predict a nearly flat geometry, with some suggesting that $\Omega_{tot} = 1.0$ to within $10^{-32}$, to avoid the Universe becoming too dense or too sparse over 10 billion years.[51] Guth proposed that an inflationary era increased the flatness of the Universe, during which the Universe was in causal contact, allowing for the equilibration of temperature and density, thus solving the horizon problem.[52] However, the predictive power of inflation has been debated.[53]

While inflation remains a compelling theory, the exact degree of final flatness may not be precisely predicted until a specific quantum theory provides a unique prediction. In the meantime, anthropic explanations, including the concept of multiverses, have been proposed to explain the specific inflation parameters, the resulting flatness, and the value of the cosmological constant—topics that are beyond the scope of this paper.[54]

# 4. Papers Revealing a Cosmological Constant 1984 to 1996

Between 1984 and 1996, several papers were published highlighting inconsistencies in the presumed matter-only Universe model that had no cosmological constant. These papers typically focused on one or more observed measurements, notably those summarized in the previous sections, and they included the Friedmann equation both with and without Λ. Here is a review of key papers, presented in chronological order.

These papers generally relied on new measurements of the Hubble constant, the Universe's mass density, the ages of the oldest stars, large-scale structure, and CMB fluctuations. Some papers emphasized the age discrepancy between the oldest stars and the calculated expansion age of the Universe. Others focused on cold dark matter models (CDM) that struggled to accurately reproduce the observed timing of galaxy formation and the morphology of large-scale structures.[55] Some papers also included the various arguments for a flat geometry that would have $\Omega_{tot} \approx 1$.

### *Vittorio and Silk (1985), Kofman and Starobinskii (1985), and Tayler (1986)*

Uson and Wilkinson measured a clear upper limit on the root-mean-square fluctuations of temperature at an angular scale of 4 arcmin, finding a fractional fluctuation $< 2.1 \times 10^{-5}$ against the 2.7 K temperature of the CMB radiation.[56] In their interpretation of this extraordinarily low upper limit they made no mention of inflation, a flat Universe, nor of the cosmological constant. This paper demonstrates that a cosmological constant was excluded from standard cosmology at the time.

Far-reaching analysis was performed both by Kofman and Starobinskii and by Vittorio and Silk who emphasized the first measurements of the temperature fluctuations (anisotropy) of the CMB.[57] The small anisotropy amplitudes on large (10°) and small scales offered support for early inflation and a flat Universe, $\Omega_M + \Omega_\Lambda = 1$.

Kofman and Starobinskii wrote in their abstract:

> Calculations of the large-scale microwave-background temperature anisotropy induced by flat-spectrum primordial adiabatic perturbations in a flat Friedmann Universe with a nonzero cosmological constant, $\Lambda$, show that the $\Lambda$-term helps overcome the difficulties posed by models with [only] cold, weakly interacting particles.[58]

Later in the same paper, they wrote:

> On the whole, then, a cosmology with cold particles but $\Lambda \neq 0$ is a decided improvement over a similar model with $\Omega_M = 1$ and $\Lambda = 0$, because it serves: 1) to make the Universe older; 2) to diminish the peculiar velocities of galaxies …

Vittorio and Silk wrote:

> However, a cold dark matter-dominated model may be consistent with the observations if $\Omega_M h > 0.05$ and $\Omega_\Lambda = 1 - \Omega_M$. Such a scheme might reconcile the astronomical determinations of $\Omega_M$ with the inflationary prediction of a flat Universe.[59]

Here, $h$ is the Hubble constant in units of 100 km/s/Mpc. The quote above corresponds to the modern model of the Universe. The wording of "may be" and "might reconcile" appropriately express the uncertainties of the day.

Tayler attempted to reconcile the expansion age of the Universe computed from the Friedmann equation with measurements of the ages of the oldest stars, along with $\rho_M$, and $H_0$, with special attention to their uncertainties.[60] He presented graphical diagnostics of models having $\Lambda = 0$ allowing visual inspection of inconsistencies. Unfortunately, the uncertainties of the three measured quantities were roughly a factor of 2, blurring his assessment of the viability of a $\Lambda = 0$ Universe.

However, Tayler noticed that models with $\Lambda = 0$ yielded an inconsistency with the contemporary estimates of the ages of globular clusters being ~15 Gyr. He wrote,

> Unless the globular clusters are very much younger than any of the recent calculations suggest, $H_0$ must be significantly less than 75, if the standard big bang model [$\Lambda = 0$] is to be valid.[61]

Tayler also showed that self-consistent solutions for $\Omega_M = 1$ and $\Lambda = 0$ were already ruled out by the observations. Tayler's analysis was seven years ahead of the needed better measurements.

### *Efstathiou, Sutherland, and Maddox (1990) and Lahav et al. (1991)*
Efstathiou, Sutherland, and Maddox argued that a cosmological constant was needed to explain the power in large scale structures.[62] The abstract is worth repeating here, to provide historical accuracy and to set the initial status for the change during the next five years:

> We argue here that the successes of the [Cold Dark Matter (CDM)] theory can be retained and the new observations accommodated in a spatially flat cosmology in which as much as 80% of the critical density is provided by a positive cosmological constant, which is dynamically equivalent to endowing the vacuum with a non-zero energy density. In such a universe, expansion was dominated by CDM until a recent epoch, but is now governed by the cosmological constant. As well as explaining large-scale structure, a cosmological constant can account for the lack of fluctuations in the microwave background and the large number of certain kinds of objects found at high redshift.[63]

This statement from Efstathiou *et al.* contains insightful reasoning and is extraordinarily prescient about the self-consistency accomplished by invoking a cosmological constant. It is difficult to imagine a more accurate rendering of the modern view.

Lahav *et al.*, summarize the reasons a cosmological constant may solve multiple issues of cosmology including the discrepancy between the age of the oldest stars and the Hubble expansion rate, consistency between the measured low mass density and both inflation and flatness, the number counts of galaxies at redshifts > 1, and difficulties of cold dark matter alone in explaining the large-scale structure observed.[64] After treating the dynamical growth of structure that fails with a matter-only flat Universe, Lahav, *et al.*, conclude,

> An alternative is to assume that all the matter in the Universe is baryonic, but to add a cosmological constant in order to save inflation (i.e. zero curvature), so … $\Omega_M + \Omega_\Lambda = 1$.[65]

### *Carroll, Press, and Turner (1992)*
Carroll, Press, and Turner summarized the state-of-the-art in an *Annual Review of Astronomy and Astrophysics* titled, "The Cosmological Constant", offering a review of all work in the field.[66] The first half of the paper describes the possible origin of the cosmological constant from quantum fluctuations in the vacuum of the Universe. Their Section 2 is titled "Why a Cosmological Constant Seems Inevitable" in which they summarize,

> For physicists, then, the cosmological constant problem is this: There are independent contributions to the vacuum energy density from the virtual fluctuations of each field, from the potential energy of each field, and possibly from a bare cosmological constant itself. Each of these contributions should be much larger than the observational bound; yet, in the real world they seem to combine to be zero to an uncanny degree of accuracy. Most particle theorists take this situation as an indication that new, unknown physics must play a decisive role.[67]

Their Section 4.2 centers on the comparison of the ages of the oldest stars in globular clusters and the expansion age of the Universe from the Hubble constant for different cosmological models. They state,

> Despite these formidable difficulties, observational and theoretical, the consensus of expert opinion concerning the ages of the oldest globular cluster stars is impressive. All seem to agree that the best-fit ages are 15-18 Gyr or more."[68]

We now know this age is too high by ~3 Gyr, and this statement does not contain citations but depends on "expert opinion", leaving open the question of which "experts" were consulted. Still, it is an accurate report. Realizing the Hubble constant near 80 km/s/Mpc corresponds to a much younger expansion age ~10.5 Gyr, they reconsider, stating,

> one wishes to know the lower limit on these oldest stellar ages; unfortunately, it is not a matter of formal errors but rather of informed judgments of how far various effects and uncertainties can be pushed. The range of expert opinion clusters around 12-14 Gyr, whether based on consideration of many clusters (Sandage & Cacciari 1990, Rood 1990) or the few best studied cases such as 47 Tuc and M92 (VandenBerg et al. 1990, Pagel 1990).[69]

Their description of uncertainties about systematic errors in the stellar ages, concluding the oldest stars could be as young as 12 Gyr old, still leaves an inconsistency with a simple $\Lambda = 0$ expansion age of 10.5 Gyr (for $H_0 \approx 80$ km/s/Mpc and $\Omega_M \approx 0.2$).[70]

They adopt $\Omega_M = 0.1$ and $\Lambda = 0$, thereby stating that $H_0$ must be less than 76 km/s/Mpc to be consistent with the age of the oldest stars around 12 Gyr. They declare no major tension in the two age determinations for the case $\Lambda = 0$, stating, "is there really any problem at the moment?"[71] They quote past low values of $H_0$ near 50 km/s/Mpc from Sandage and others, so low that, "a value of $H_0$ small enough to avoid any age concordance problems, even in an $\Omega_M = 0.1$ and $\Lambda = 0$ model, is not yet excluded".[72] Later they state, "an open model consisting entirely of baryons with $\Lambda = 0$ and $\Omega_M = 0.1$ is by no means strongly excluded."[73] Thus, Carroll *et al.* remained comfortable with not invoking a cosmological constant.

In retrospect, by the time Carroll, *et al.*, was formally published in 1992, progress had already occurred in the observations, showing $\Omega_M$ to be twice as large, 0.2-0.3, and also showing the Hubble constant to be 68 to 90 km/s/Mpc as described above. So, the answer to the Carroll *et al.*'s question in 1992, "is there really any problem at the moment?" was "no, but wait two or three years."

### *Kofman, Gnedin, and Bahcall (1993)*
Kofman, Gnedin, and Bahcall conducted simulations of the growth of large-scale structure, using the newly obtained 10° angular scale anisotropy in the CMB from COBE.[74] They compared their simulation results with recent observations of galaxy power spectra, peculiar velocities, and the galaxy cluster mass function. Their findings suggested that models relying solely on cold dark matter might not adequately connect the CMB fluctuations on large scales with the current large-scale structure of the Universe unless a cosmological constant was included. They wrote, "We find that a flat model ($\Omega_m + \Omega_\Lambda = 1$) with $\Omega_m \approx 0.25\text{-}0.3$ is consistent with all the above large-scale structure observations."[75] However, they also discussed the "Pros and Cons" of invoking a cosmological constant.

### *Roukema and Yoshii (1993)*
Roukema and Yoshii extracted galaxy merger history trees from N-body simulations of galaxy mergers to compute the effect of merging on the galaxy angular two-point autocorrelation function including luminosities.[76] They initially adopted a flat model of the Universe containing matter only ($\Lambda = 0$), and they used the luminosity function of Efstathiou, Ellis and Peterson.[77] They found that the simulated two-point correlation function was significantly different from that observed. Faced with that conflict, they end their "Discussion and Conclusions" section with, "Another alternative, though perhaps somewhat premature, would be a flat universe with $\Omega_m \approx 0.2$, $\Omega_\Lambda = 1 - \Omega_m$ ... This has the additional cosmological advantage of enabling the globular cluster ages to be less than that of the universe and the interesting consequence that the deceleration parameter, $q_o = \Omega_m / 2 - \Omega_\Lambda$, would be negative, i.e., the local universe would be 'accelerating'!"[78] This is a rare statement, prior to 1998, describing the acceleration of the Universe expansion and $q_o$ being negative.

### *Krisciunas (1993)*

Krisciunas[79] compared the expansion-based age of the Universe to the age of the oldest stars using contemporary measurements, adopting $H_0 = 80 \pm 11$ km/s/Mpc as reported by Jacoby, *et al.*, and van den Bergh. To calculate the age of the Universe, he integrated the Friedmann Equation to determine the "look-back" time to the Big Bang, following the method of Carroll, *et al.* Krisciunas created diagnostic plots showing the expansion age of the Universe versus $H_0$ for different assumed values of $\Omega_M$, including 0.0, 0.03, 0.10, 0.3, and 1.0.

He then examined the estimated ages of the oldest stars in globular clusters (see previous sections here) settling on an age of $13.5 \pm 2.0$ Gyr. He adopted 1.5 Gyr as the time between the Big Bang and the formation of these stars (a period now known to be only ~0.4 Gyr) to estimate an empirical age of the Universe at $T_0 = 15.0 \pm 2.5$ Gyr. Using his diagnostic plot for $\Lambda = 0$ (his Figure 3) at $H_0 = 80 \pm 11$ km/s/Mpc, Krisciunas simply "read-off" the corresponding value of $\Omega_M$ from the provided loci. He wrote,

> If $T_0 = 15.0$ Gyr and $H_0 = 80$ km/s/Mpc, an inspection of figure 3 immediately indicates that the density of the universe must be less than zero![80]

Krisciunas then hunted for resolutions, including a lower age of the Universe of only 13.5 Gyr and $H_0 < 80$. Adopting the observed $\Omega_M \approx 0.2$, he found that self-consistent solutions are not apparent. He wrote,

> But as long as the determinations of Hubble's constant give $H_0 \approx 80$, we must take seriously the possibility that $\Lambda > 0$.[81]

He concluded,

> Sensible values of $T_0$ and $H_0$ point to a positive value for the cosmological constant.[82]

Krisciunas explicitly states that the combination of the newly measured $H_0$, the youngest plausible age of the oldest stars at 12 Gyr, and $\Omega_M \approx 0.2$ is inconsistent with $\Lambda = 0$, instead suggesting a positive cosmological constant.

Krisciunas notes that to maintain a $\Lambda = 0$ scenario, one would need to assume "one-sided" 1-$\sigma$ errors in $H_0$, the age of the oldest stars, and $\Omega_M$ – essentially that their true values differ in a way that "saves" $\Lambda = 0$. While this possibility was still viable in 1993, it was increasingly strained. Modern measurements indeed show both $H_0$ and the age of the oldest stars are slightly lower by 1$\sigma$ compared to the 1993 values, which could help support $\Lambda = 0$. However, the modern value of $\Omega_M$, around 0.3, is higher than the 0.2 previously assumed, making $\Lambda = 0$ unlikely. The following section provides a quantitative probability assessment.

### *Krauss and Turner (1995)*

Krauss and Turner considered a wide range of contemporary cosmological measurements, including $H_0$, the age of the oldest stars, mass density, large-scale structure, and CMB fluctuations on a 10° angular scale.[83] They incorporated these factors into the Friedmann equation and evaluated the simulated growth of structure. They also considered the "theoretically favored flat Universe." They concluded, as stated in the first sentence of their abstract,

> A diverse set of observations now compellingly suggest that the Universe possesses a nonzero cosmological constant.[84]

They highlight that the improved value of $H_0 = 80 \pm 5$ km/s/Mpc is supported by early results from the Hubble Space Telescope study of Cepheid variables. They also adopt the refined estimate for the age of the oldest stars, which provides a 1-sigma lower limit of 13 Gyr. This lower limit precludes $\Lambda = 0$ in the context of an assumed flat Universe.

Krauss and Turner summarize this age issue:

> The age problem is more acute than ever before. A cosmological constant helps because for a given matter content and Hubble constant the expansion age is larger. For example, for a flat universe with $\Omega_M = 0.2$ and $\Omega_\Lambda = 0.8$, the expansion age is 1.1 $H_0^{-1}$ = 13.2 Gyr for a Hubble constant of 80 km/s/Mpc.[85]

Their statements foresaw the development of the modern-day $\Lambda$CDM model, which resolves the age discrepancy – three years before the Type Ia supernova results were published.

Krauss and Turner further emphasize that the measured 10° angular scale spatial fluctuations of the CMB[86] self-consistently seed the growth of the observed large-scale structure if a cosmological constant is invoked that renders the Universe flat.[87] Krauss and Turner emphasize that all measurements are consistent with a flat Universe having $\Omega_M \sim 0.2$ and $\Omega_\Lambda \sim 0.8$.

Krauss and Turner summarize their conclusions:

> Cosmological observations thus together imply that the "best-fit" model consists of matter accounting for 30%-40% of critical density, a cosmological constant accounting for around 60%-70% of critical density, and a Hubble constant of 70-80 km/s/Mpc. We emphasize that we are driven to this solution by simultaneously satisfying a number of independent constraints.[88]

Their considerations depend on the theoretical arguments for a flat Universe. They conclude,

> Unless at least two of the fundamental observations described here are incorrect a cosmological constant is required by the data.[89]

*Ostriker and Steinhardt (1995)*

Ostriker and Steinhardt wrote two short papers that summarize the evidence for a cosmological constant.[90] These two papers adopt the same new measurements of $H_0 = 80$, a lower limit $\Omega_M > 0.2$, and ages for the oldest stars of 13 to 16 Gyr, as reviewed here, similar to Krauss and Turner.[91] The Ostriker and Steinhardt papers emphasize the implications of inflation, the isotropy and limits on fluctuations of the CMB from COBE, and the models that simulate large-scale structure. They emphasize limits on $\Omega_\Lambda < 0.75$ from gravitational lensing, which merited later consideration.

Ostriker and Steinhardt submitted one paper only to the online arXiv (not refereed), and it contains this important statement in the abstract,

> For these models, microwave background anisotropy, large-scale structure measurements, direct measurements of the Hubble constant, $H_0$, and the closure parameter, $\Omega_M$, ages of stars and a host of more minor facts are all consistent with a spatially flat model having significant cosmological constant, $\Omega_\Lambda = 0.65 \pm 0.1$, $\Omega_M = 1-\Omega_\Lambda$ (in the form of 'cold dark matter').[92]

They identify the "concordance domain" of parameter space that meets all the observational constraints, concluding in their published paper:

> The observations do not yet rule out the possibility that we live in an ever-expanding "open" Universe, but a Universe having the critical energy density and a large cosmological constant appears to be favoured.[93]

*Bagla, Padmanabhan, Narlikar (1996)*

Bagla et al.[101] employed updated measured values of the Hubble constant, including one newly derived from Cepheids with the Hubble Space Telescope and one from new SNe Ia measurements. They also included the latest measured total mass density of the universe and the ages of globular clusters, all with associated uncertainties. They carefully considered models both with and without a cosmological constant. They showed clearly that the only surviving models, given the uncertainties, were models that included a large cosmological constant. They wrote in their conclusions:

"Even allowing for errors on both fronts [Hubble constant and stellar evolution], the conclusion today is inescapable that the standard big bang models *without* the cosmological constant are effectively ruled out." (italics theirs.)

"By the same token, we would have insisted that $\Omega_\Lambda = 0$. Such a model is clearly ruled out by the observations. It is indeed hard to understand why the leftover cosmological constant is such as to exactly conform to the flatness condition."

# 5. The Probability Prior to 1998 that $\Lambda = 0$

Before 1998, the prevailing view largely disregarded the cosmological constant. However, the handful of papers reviewed in the previous section insightfully assembled the latest measurements of the Universe, exposing internal inconsistencies with that prevailing view. Those new measurements, along with their associated uncertainties, directly projected to different values of Λ, including the probability that Λ = 0.

The previous sections of this paper summarize the newly measured ages of the oldest globular clusters by Chaboyer, *et al.*, VandenBerg, Bolte, and Stetson, and by Bolte and Hogan.[94] These studies accounted for uncertainties in the input stellar physics, such as opacities, nuclear cross sections, and convection. Chaboyer *et al.*'s Monte Carlo trials indicated a 95% probability that the oldest globular clusters were older than 12.1 Gyr.[95] The other studies agreed, estimating the oldest stars' ages at $15.8 \pm 2.1$ Gyr, with an approximate 2-σ lower limit of 11.6 Gyr.

A one-tailed probability estimate is appropriate here, as only the lowest possible age is relevant. Using the lower stellar age estimate of 11.6 Gyr (which leans towards compatibility with Λ = 0), there is about a 2.5% probability that the oldest globular clusters are younger than 11.6 Gyr. (Modern measurements place their age at 12.5 Gyr – see next section.) Given that stars require about 0.4 Gyr to form after the Big Bang (from the highest known galaxy redshifts of 14, from the James Web Space Telescope), there was roughly a 2.5% chance that the Universe was younger than 12 Gyr. The ages of the oldest white dwarfs (described here previously) were in agreement. *Therefore, in 1996, the measured ages of the oldest stars suggested only a ~2.5% probability that the Universe was less than 12 Gyr old.*

To use contemporaneous values for the early 1990s, the most distant galaxies known had a redshift of 4.9.[96] That implies they formed 1.1 Gyr after the Big Bang. Adding that larger duration, 1.1 Gyr, to the age of the earliest stars implies a 2.5% probability that the Universe is less than 12.5 Gyr old, i.e. even older. Indeed, modern estimates of the Universe's age are near 13.8 Gyr (see next section).

Importantly, this requirement from the oldest stars that the Universe must be older than 12 Gyr can be compared to the calculated expansion-age of the Universe assuming Λ = 0. In 1996, the measured parameters were $H_0 = 80 \pm 11$ km s$^{-1}$ Mpc$^{-1}$ and $\Omega_M = 0.25 \pm 0.1$ (e.g., Bahcall, Fan, and Cen).[97] Assuming Λ = 0, the resulting expansion age of the Universe is 10.1 Gyr, which is obviously inconsistent with the age, 12 Gyr, from the oldest globular clusters. To address this conflict, one could adopt the 1-σ lower limit of $H_0 = 69$ km s$^{-1}$ Mpc$^{-1}$. This gives an expansion age of 11.7 Gyr - still in conflict with the 2-σ lowest possible age of the Universe of 12 Gyr from stars. *This indicates a formal 2.5% probability that Λ = 0 was correct, as of the mid-1990s.*

The tension between the highest possible expansion age of 11.7 Gyr (with $H_0 = 69$, $\Omega_M = 0.25$, and Λ = 0) and the 2-σ lower-end age of the Universe ~12 Gyr, based on the oldest stars, indicates a low probability of consistency. Adopting modern values of $H_0 = 72$, $\Omega_M = 0.3$, while keeping Λ = 0, increases the implied expansion age to 11.0 Gyr - still lower than the age of the Universe from globular clusters of ~12 Gyr.

This point deserves emphasis. Using the improved mass density measurements by Bahcall, Fan and Cen of $\Omega_M = 0.3 \pm 0.1$, which agrees with the modern value, and using a modern Hubble constant of $H_0 = 72$ km s$^{-1}$ Mpc$^{-1}$, leads to an expansion age of the Universe of 11.0 Gyr assuming Λ = 0.[98] This expansion age is inconsistent with the lower bound of 12 Gyr derived from the oldest stars, reinforcing the low probability of about 2.5% of Λ = 0 being correct.

In summary, considering the uncertainties in the measurements, any proposition that $\Lambda = 0$ carried a probability of less than 2.5% in 1996. This low probability holds true whether one adopts the nominal measurements from 1996 for the Hubble constant and mass density or instead adopts adjusted measurements, given their uncertainties, to purposely attempt to accommodate $\Lambda = 0$. The calculated expansion age, the time since the Big Bang given the observed expansion rate and assuming $\Lambda = 0$, is simply too young to accommodate the older age of the oldest stars. *The likelihood of achieving consistency across all measurements with $\Lambda = 0$ was no more than a few percent in 1996.*

It is common to hear astronomers dismiss the ages of the oldest stars, claiming the uncertainties were too large to take seriously. However, this overlooks the fact that globular cluster ages were 2-$\sigma$ to 3-$\sigma$ *larger* (i.e., several billion years older) than the expansion age of ~10.5 Gyr if one simply assumed $\Lambda = 0$. The discrepancy was large. Similarly, the missing neutrino problem from the Sun was often dismissed as merely due to poor stellar interior models. However, both issues rely on the same underlying stellar interior physics.[99] Ignoring the ages of the oldest stars is reminiscent of those who dismissed the Standard Solar Model, which was also vindicated with the discovery of neutrino oscillations.

The 2.5% probability that $\Lambda = 0$ was correct around 1996 should be adjusted downward below 2.5%, as it doesn't account for other indications that $\Lambda = 0$ was problematic. Theoretical considerations of the horizon problem and early inflation pointed to a flat Universe with $\Omega_\Lambda \sim 0.7$ as described previously here. Also, simulations of the growth of large-scale structure successfully matched observations by employing both $\Lambda = 0$ and $\Omega_\Lambda \sim 0.7$, showing equal probability. Only a prior bias would favor any value. Considering these factors, by 1996 the probability of $\Lambda = 0$ was likely less than 2.5%.

The pre-1998 specific prediction of $\Omega_\Lambda = 0.7$ by Krauss & Turner and by Ostriker & Steinhardt cannot be dismissed as a stroke of casino luck.[100] Imagine them at a roulette table, where the spinning wheel is marked with many possible values of $\Omega_\Lambda$, with a green "Zero" occupying up a big slice of the wheel. They pushed their chips on $\Omega_\Lambda = 0.7$. When the ball finally settled, not only were they right about $\Omega_\Lambda$ being non-zero, but they also predicted its value.

Independently the low probability for $\Lambda = 0$ was published by others. Bagla, Padmanabhan, and Narlikar in 1996 support this low probability in their conclusions, "Even allowing for errors on both fronts, the conclusion today is inescapable that the standard big bang models *without* the cosmological constant are effectively ruled out."[101] Similarly, leaning on the need for inflation, Turner and White write, "This dilemma can be resolved if a smooth component contributes the remaining energy density ($\Omega_X = 1 - \Omega_M$)."[102] Kennicutt, Freedman, and Mould wrote in 1995, "if the uncertainties attached to these $H_0$ and age values prove to be conservative, then *no combination of the parameters* is consistent with a $\Lambda=0$ cosmology."[103] The low probability that $\Lambda$ was simply zero was appreciated prior to 1996.

# 6. Modern Measurements Applied to Pre-1998 Reasoning

The previous section presented a calculation of the probability that $\Lambda = 0$, based on the measurements known in 1996. Modern measurements clarify the integrity of those predictions.

**Modern Ages of the Oldest Globular Clusters**

Hipparcos provided revised distances to key photometric benchmarks, which in turn adjusted the distances to globular clusters. Chaboyer *et al.* showed that these adjustments significantly reduced the estimated ages of globular clusters.[104] They noted, "Hipparcos parallax measurements suggest a large systematic shift of approximately 0.2 mag compared to earlier estimates. This has the effect of reducing the mean age of the oldest globular clusters by almost 3 Gyr."[105] They further stated, "Our best estimate for the mean age of the oldest globular clusters is now $11.5 \pm 1.3$ Gyr."[106]

VandenBerg *et al.* provided absolute ages and uncertainties for 55 globular clusters (GCs) using Hubble Space Telescope photometry and isochrones to fit the main-sequence turn-off.[107] They found that the oldest GCs have ages of $13 \pm 1.5$ Gyr (see their Figure 33).

More accurate, modern age determinations are based on even more precise GAIA-based parallaxes and improved stellar interior physics. Thompson *et al.* determined that 47 Tuc has an age of $12.0 \pm 0.5$ Gyr.[108] Massari, Koppelman, and Helmi found that the most metal-poor globular clusters have ages of $12.5 \pm 0.6$ Gyr (see their Figures 1 and 4).[109]

Don VandenBerg used the horizontal branch of clusters, particularly in the rest-frame near IR, finding that using horizontal branch photometry instead of the main-sequence turn-off yields nearly the same ages for the oldest globular clusters, ranging from 11.8 to 12.8 Gyr. Thus, the oldest globular clusters have ages of ~$12.3 \pm 0.6$ Gyr. The oldest galaxies known now (from the James Web Space Telescope) have a redshift of 14, implying stars formed 0.3 Gyr after the Big Bang. Adding ~0.3 Gyr for the minimum time to form globular clusters after the Big Bang, *the revised age of the Universe, based on globular clusters alone, is at least $12.7 \pm 0.6$ Gyr.* In comparison, for $\Lambda = 0$ and modern values of $H_o = 70$ and $\Omega_M = 0.3$ (which are not significantly different from those in 1996), the calculated expansion age of the Universe is only 11.3 Gyr.

Modern ages of globular clusters imply the age of the Universe is at least 12.7 Gyr, which still exceeds the expansion age of 11.3 Gyr if $\Lambda = 0$. This demonstrates that the pre-1998 papers that indicated a non-zero cosmological constant were grounded in both sound logic and adequately accurate measurements, not contradicted by modern measurements. It's remarkable that the (painstakingly obtained) observations available in 1996—before the use of supernovae and 0.5° CMB anisotropy measurements—already strongly argued against $\Lambda = 0$, even when compared with today's more precise values.

Clearly, a flat Universe composed of mass only was ruled out. Chaboyer *et al.* state, "First and foremost, this result suggests that the long-standing conflict between the Hubble and GC age estimates for a flat matter-dominated universe is now resolved for a realistic range of Hubble constants."[110] Indeed, a *flat, matter-dominated* Universe was ruled out. But an open, matter-dominated Universe was also unlikely, as the expansion age was too short compared to the globular cluster ages.

Turner and White explicitly noted that models with a cosmological constant were favored, stating, "Though $\Lambda$CDM is the 'best fit' CDM model the theoretical motivation is weak", referring to the required tiny vacuum energy.[111]   Turner and White also note evidence against ΛCDM from the magnitude-redshift (Hubble) diagram of Type Ia supernovae, as Perlmutter *et al.* reported $\Omega_\Lambda = 0.06^{+0.28}_{-0.34}$.[112]

**Large Scale Structure and CMB**
Several observations of large-scale structure, notably from the CfA survey and the Lick galaxy survey, were instrumental in determining the structure of the baryonic and dark matter of the present Universe. Thanks are due to Steve Shectman, Marc Davis, Margaret Geller, and John Huchra for leading these surveys. These large-scale surveys were monumental efforts involving both angular and redshift measurements. After 1996, the work of Navarro, Frenk, and White became particularly important as they simulated dark matter halos in hierarchically clustering universes. They noted, "We find that all such profiles have the same shape, independent of the halo mass, the initial density fluctuation spectrum, and the values of the cosmological parameters."[113] Their simulations did not reveal differences in large-scale structure among models with or without a cosmological constant, leaving both $\Lambda = 0$ and $\Omega_\Lambda = 0.7$ as plausible scenarios for large-scale structure.

Similarly, CMB data around 1995 did not definitively support either a flat or open Universe.[114] The concept of a "flat" geometry based on "inflation" remained theoretically compelling but was not yet backed by unique observational predictions. It was only in 2000, with the measurement of the 0.5-degree angular scale of the CMB by COBE and other CMB experiments, that the flat Universe was empirically supported.

**Λ = 0 has Low Probability with Modern Observations**
The simplest evidence against $\Lambda = 0$ stemmed from the age discrepancy, which emerged in 1995 and holds even with modern 2024 measurements.  Modern Universe measurements show that $H_0 = 70 \pm 3$ km s$^{-1}$ Mpc$^{-1}$ and $\Omega_M = 0.30 \pm 0.01$, implying an expansion age of the Universe (assuming $\Lambda = 0$) of 11.25 Gyr.  In contrast, the age of oldest globular clusters is 12.3 +- 0.6 Gyr, and they formed at least 0.2 Gyr after the Big Bang. Thus, the oldest stars show the Universe is at least 12.5 Gyr old.  *A model of cosmic expansion that ignores acceleration from dark energy implies the universe is only 11.25 billion years old, which is not possible given its minimum age of 12.5 billion years derived from the oldest stars.* The model is not favored.

The reasoning and the evidence against $\Lambda = 0$ remain as valid with modern measurements as they were in 1996. In fact, the age discrepancy that arises from assuming $\Lambda = 0$ has only become more pronounced with modern measurements.

# 7.  CONCLUSIONS

Between 1990 and 1995, observational evidence for a significant cosmological constant, now known as "dark energy," emerged and strengthened, and it arrived several years before the powerful supernova evidence in 1998. By 1995, new measurements—including the Hubble constant, matter density, ages of the oldest stars, large-scale structure, small CMB fluctuations, and inflationary predictions of a flat Universe—created significant tension for models that simply assumed $\Lambda = 0$.

This review includes a formal calculation of the probability that the Universe contains no dark energy ($\Lambda = 0$) based on the measurements available in 1995. The discrepancy between the age of the Universe from globular clusters and the computed expansion history left the probability of $\Lambda = 0$ at less than 2.5%. Additionally, simulations of large-scale structure and arguments for a flat Universe further reduced this likelihood.

Five papers (and several others cited herein) clearly stated the value of invoking a non-zero cosmological constant:

- Efstathiou *et al.* and Lahav *et al.* noted clearly that the age discrepancy and the formation of large-scale structure could be self-consistently explained by invoking a cosmological constant.[115]

- Krisciunas stated, "Sensible values of $T_0$ and $H_0$ point to a positive value for the cosmological constant."[116]

- Krauss and Turner emphasized, "unless at least two of the fundamental observations described here are incorrect, a cosmological constant is required by the data."[117]

- Ostriker and Steinhardt stated, "a large cosmological constant appears to be favoured."[118]

Several other papers published between 1985 and 1997, notably those by Bahcall *et al.* and Bagla, Padmanabhan, and Narlikar, describe the growing viability of models with a nonzero cosmological constant.[119]

It is remarkable that such a profound and mysterious phenomenon as dark energy could be inferred through the meticulous work of hundreds of scientists across various subfields, using different instruments, and making measurements at the cutting edge of modern technology. Between 1991 and 1995, the combination of this vast array of data—including the assessments of both random and systematic uncertainties—and standard general relativity, applied on scales of billions of light-years, led to a self-consistent model that pointed to the existence of dark energy.

Surprisingly, many of these forward-looking papers from this period are not widely cited, and none are currently mentioned in Wikipedia entries on the cosmological constant. This lack of recognition for the early discovery of dark energy could be due to several factors. The cosmological constant itself had a chequered history, and even today the physics community struggles to explain it through virtual quantum fluctuations. Additionally, the rapidity of the improvement of various measurements in the early 1990s may have contributed to the hesitance in acknowledging these early findings. Intellectual inertia also may have played a role, as the scientific community is often slow to abandon paradigms.[120]

Interestingly, the supernova papers by Perlmutter, Goldhaber, Pennypacker, and Goobar in the 1990s rarely mentioned $\Lambda$ until May 1995, which was published one month after Krauss and Turner.[121] This timing raises the question of whether papers (and personal discussions) about the evidence for $\Lambda$ may have emboldened the two Type Ia supernova groups to reconsider $\Lambda$.

The credibility of dark energy gained from Type Ia supernova observations was bolstered by the direct and visually understandable nature of the measurements.[122] These brilliant observations provided clear evidence of accelerating expansion, with distances at higher redshifts being greater than expected, making the interpretation straightforward. In contrast, the earlier arguments for dark energy before 1998 required a more nuanced understanding of the new data and the magnitude of the errors involved.

The work of M. Turner, L. Krauss, G. Efstathiou, O. Lahav, L. Kofman, N. Bahcall, B. Roukema, Y.Yoshii, K. Krisciunas, S. Carroll, W. Press, E. Turner, J. Ostriker, P. Steinhardt, J. Bagla, T.Padmanabhan, J.V. Narlikar, and others revealed inconsistencies between the impressive new observations and the standard model of an expanding Universe. Credit must go to the countless observational astronomers who worked tirelessly to accurately measure the Hubble constant, mass density, large-scale structure, stellar ages, and CMB anisotropy.

This collective effort did more than reveal "tension"; in hindsight, it showed that the probability of $\Lambda = 0$ was only a few percent. They achieved a concordance of all observational constraints only by invoking a significant positive value for $\Lambda$. The theorists and observers identified a "best-fit model" with $\Omega_M$ = 0.3-0.4 and $\Omega_\Lambda$ = 0.6-0.7—a model that has not only endured but has also been strengthened over time. This community of dedicated scientists demonstrated the extraordinary power of the modern scientific method on a grand scale.

## 7. ACKNOWLEDGEMENTS

This work benefitted from valuable communications with Alex Filippenko, Neta Bahcall, Michael Turner, Kevin Krisciunas, Lawrence Krauss, Susan Kegley, John Gertz, Andrew Fraknoi, Brian Hill, and Gibor Basri. We thank Ned Wright for his "Cosmology Calculator". We also thank Space Laser Awareness for their technical support.

### DATA AVAILABILITY

This paper contains no new data. The statistical analysis is based on existing, published data.

### NOTE ON CONTRIBUTOR

Geoff Marcy led the discovery of 70 of the first 100 exoplanets ever found, including the first multi-planet system orbiting a Sun-like star. He was awarded the Shaw Prize with Michel Mayor in 2005. In 2013, Erik Petigura, Andrew Howard and Marcy used NASA *Kepler* data to discover that 20% of Sun-like stars harbor a warm, Earth-size planet, implying tens of billions of habitable worlds within our Milky Way. Marcy is director of the non-profit Center for Space Laser Awareness.

[118] Ostriker and Steinhardt, "The observational case for a low-density universe with a non-zero cosmological constant" (ref. 90), p. 600.

[119] Bahcall, Fan, and Cen, "Constraining $\Omega$ with cluster evolution" (ref. 36). Bagla, Padmanabhan, and Narlikar, "Crisis in cosmology: observational constraints on Omega and $H_0$" (ref. 101).

[120] T. Kuhn, *The Structure of Scientific Revolutions* (Chicago: University of Chicago Press, 1962).

[121] A. Goobar and S. Perlmutter, "Feasibility of measuring the cosmological constant lambda and mass density Omega using type Ia supernovae", *The Astrophysical Journal*, 450 (1995), 14-18. Krauss and Turner, "The Cosmological Constant Is Back" (ref. 17).

[122] A.G. Riess, A.V. Filippenko, P. Challis, *et al.*, "Observational evidence from supernovae for an accelerating universe and a cosmological constant", *The Astronomical Journal*, 116 (1998), 1009–38. S. Perlmutter, G. Aldering, G. Goldhaber, *et al.*, "Measurements of $\Omega$ and $\Lambda$ from 42 high-redshift supernovae", *The Astrophysical Journal*, 517 (1999), 565–86. A.V. Filippenko, "Einstein's biggest blunder? high-redshift supernovae and the accelerating universe", *Publications of the Astronomical Society of the Pacific*, 113 (2001), 1441–8.